\documentclass[twocolumn,amsmath,amssymb,aps,floatfix,showkeys]{revtex4-2}

\usepackage[caption=false]{subfig}
\usepackage{graphicx}
\usepackage[margin=1.5cm]{geometry}

\usepackage{dcolumn}
\usepackage{bm}
\usepackage{tabularx}
\usepackage{multirow} 
\usepackage{booktabs}
\usepackage[usenames,dvipsnames,svgnames,table]{xcolor}
\usepackage{hyperref}
\hypersetup{
	colorlinks=true,
	linktocpage=true,
	citecolor=YellowOrange,
	linkcolor=violet,
	urlcolor=MidnightBlue!50!Aquamarine,
}
\usepackage[range-phrase=-, range-units=single, separate-uncertainty=true]{siunitx}
\DeclareSIUnit\century{century}
\DeclareSIUnit\year{year}
\DeclareSIUnit\day{day}
\DeclareSIUnit\parsec{pc}

\usepackage{tikz}

\begin{document}

\preprint{APS/123-QED}

\title{On a new statistical technique for the real-time recognition of ultra-low multiplicity astrophysical neutrino burst}

\author{Marco Mattiazzi}
\author{Mathieu Lamoureux}
\author{Gianmaria Collazuol}

\affiliation{Dipartimento di Fisica, INFN Sezione di Padova and Università di Padova, I-35131, Padova, Italy}

\date{\today}

\begin{abstract}
The real-time recognition of neutrino signals from astrophysical objects with very-low false alarm rate and short-latency, is crucial to perform multi-messenger detection, especially in the case of distant core-collapse supernovae accessible with the next generation of large-scale neutrino telescopes. The current time-based selection algorithms implemented in operating online monitors depend mainly on the number of events (multiplicity) detected in a fixed time window, under the hypothesis of Poisson-distributed  background. However, these methods are not capable of exploiting the time profile discrepancies between the expected supernova neutrino burst and the stationary background.

In this paper we propose a new general and flexible technique (beta filter method) which provides specific decision boundaries on the cluster multiplicity-duration plane, guaranteeing the desired false alarm rate in an analytical way. The performance is evaluated using the injection of a general purpose SN-like signal on top of realistic background rates in current detectors. An absolute gain in efficiency of up to $\sim 80\%$ is achieved compared with the standard techniques, and a new ultra-low multiplicity region is unveiled.
\end{abstract}

\maketitle

\section{Introduction}
The definition of the neutrino burst recognition algorithm is the key aspect for any real-time astrophysical neutrino monitor e.g., the one looking for prompt signals from core-collapse supernovae (CCSNe). Several search strategies have been implemented and proposed by different large-scale experiments  along the years~\cite{Abe2016,Agafonova2008,Fulgione1996}. Nevertheless, all of them need the definition of a clustering procedure before making the signal-to-noise discrimination.

\begin{figure}[!ht]
	\includegraphics[width=\columnwidth, height = 5cm,keepaspectratio]{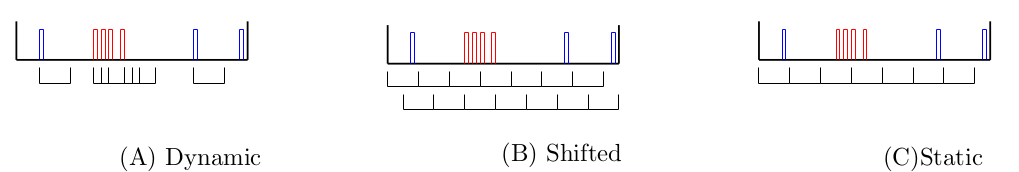}
	\caption{Schematic of the dynamic, shifted and static clustering. The red bars indicate the candidate signals, whereas the uncorrelated background is in blue. In the shifted case the first and second scans are 
shown.}
	\label{fig:clustering_scheme}
\end{figure}

Let $w$ and $T$ be, respectively,  the time window size and  the total observation time,
three possible online data clustering are displayed in the~\autoref{fig:clustering_scheme} and are henceforth referred to as:

\begin{itemize}
\item static clustering:  the time interval $T$ is divided into $N$  sliding time-windows (size $w$);
\item shifted clustering: a first scan is performed as the static case, but, with an additional scan,
starting from the middle $(w/2)$ of the previous time window;
\item dynamic clustering: each event is considered as the starting point of the time window.
\end{itemize}
The dynamic clustering preserves the entire timing information of the signal within the time window $w$ although the computational complexity grows with increasing background rates. Instead, the substantial bias of the static case is circumvented in the shifted clustering without introducing significant latency, albeit some information is still missed. Hence, we shall consider both the dynamic and shifted cases in this paper.

The typical time window size $w$ is set to \SI{20}{\second}, as this is the expected time scale of the CCSN neutrino emission during the cooling phase, consistent with the only experimental data \cite{Alekseev:1988gp,Bionta:1987qt,VanDerVelde:1987hh} available so far.

In general, the basic features of a real-time cluster are:
\begin{itemize}
\item the multiplicity, i.e. the number of events within the time window;
\item the time difference between the first and the last event in the cluster.
\end{itemize}

The standard method to evaluate online the statistical significance of a candidate cluster is based on the so-called \emph{imitation frequency} \cite{Agafonova2008,Casentini2018,SNEWS2004}.
Assuming a Poisson-distributed background with constant rate $r$, the imitation frequency
$\textrm{F}_{im}$, for a cluster with multiplicity $m$,
is 

\begin{equation}
\textrm{F}_{im}(m \mid r,w) = N_{\rm windows} \times \sum_{k=m}^{\infty}
\frac{(rw)^{k}  e^{-rw}}{ k!},
\label{eq:imitation_frequency}
\end{equation}
where $N_{\rm windows}$ is the number of time windows with fixed size $w$ in a given false alarm time ($t_{FAT}$), which number, in turn, depends on the clustering algorithm. In the first iteration of SNEWS~\cite{SNEWS2004}, the false alarm rate has been set to one false positive per century for a stand-alone operating monitor, but it could be greatly reduced if it belongs to a network of large-scale neutrino telescopes as stressed in \cite{SNEWS2021}.

In this method, once the time window size is established, the signal-to-noise discrimination depends only on the cluster multiplicity threshold ($\hat{M}$). As pointed out in \cite{Casentini2018, Halim:2019esz, lamoureux2021identification} the timing information of the burst is not fully exploited.

\begin{figure}[!ht]
	\includegraphics[width=\columnwidth, height = 8cm,keepaspectratio]{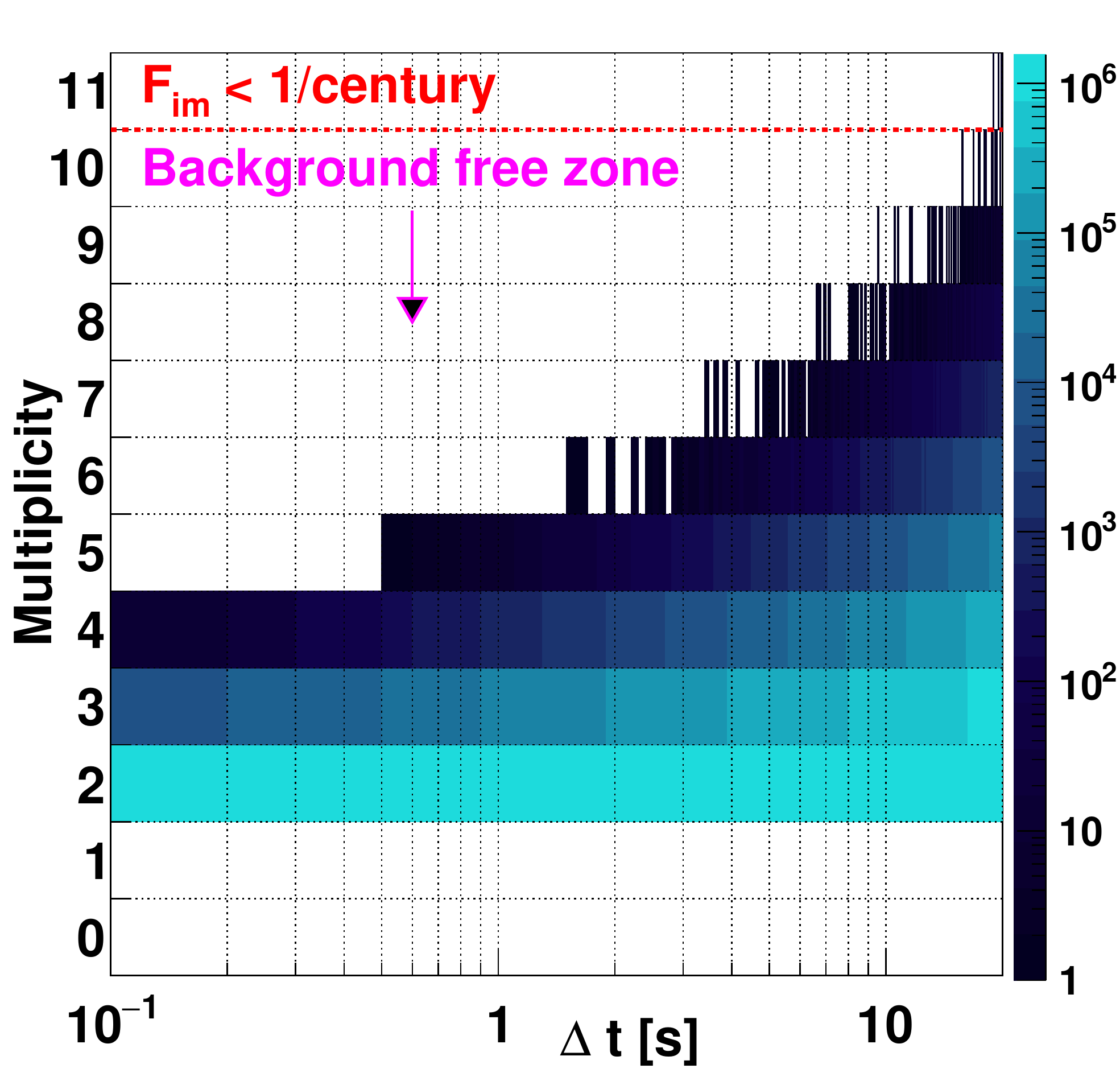}
	 \caption{ Simulated bi-variate distribution with respect to  multiplicity ($m$) and duration ($\Delta t$) of the Poisson-based (rate $r = 3 \times 10^{-2}$ Hz) background clusters performing the dynamic clustering. The simulation time is  10 times larger than the required false alarm time ($t_{FAT} = 1$ century). The dashed red  line represents the standard imitation frequency threshold as reported in~\cite{Fulgione1996,Casentini2018,lamoureux2021identification} and the violet arrow indicates the background free region beneath it, which can be exploited with the new methods described in ~\autoref{sec:betafilter}.  }
  \label{fig:bivariate}
\end{figure}

Such inefficiency is illustrated in \autoref{fig:bivariate}, where the Poisson-distributed background events have been simulated for ten centuries with a constant rate $r = \SI{3e-2}{\hertz}$ and performing the dynamic clustering. As a matter of fact, the potential low-multiplicity signal clusters, falling into the background-free zone, are below the fixed threshold defined by the imitation frequency (assuming $t_{FAT} = 1$ century), even though it does not contain any background events over ten times the required false alarm time.

Exploiting the duration information is thus clearly promising to improve the selection of low-multiplicity signal clusters.

\section{Beta filter Method}
\label{sec:betafilter}

The probability to observe at least $m$ events in a given time interval $[0,t]$, under the hypothesis of stationary Poisson process with constant rate $r$, is provided by the survival function of the Poisson distribution, also known as the cumulative of the Erlang distribution:
\begin{equation}
    F_{\gamma}(m \mid r,t) = 1 - \sum_{k=0}^{m-1} \frac{(rt)^{k} \, e^{-rt}}{k!}  = \sum_{k=m}^{\infty} {\rm{Pois}}\, (k, rt).
\label{eq:cumulativeGamma}
\end{equation}

However, in a fixed sliding time window $W = [0, w]$, the $m$ events are uniformly distributed  within $W$ according to $0<t_1<t_2<\ldots<t_m<w$. Hence, after normalizing to [0,1] by defining $x_k := t_k/w$, the normalized time of the k-th event follows the k-th order statistics from the uniform distribution:
\begin{equation}
    f_{\beta}(x_k \mid m) = \textrm{Beta}(x_k; k, m+1-k),
\label{eq:BetaNormalizedTime}
\end{equation}
and transforming back in the time domain, we get:
\begin{equation}
    f_{\beta}(t_k \mid m,w) = \dfrac{1}{w} \times \textrm{Beta}\left( \frac{t_k}{w}; k, m+1-k \right).
    \label{eq:BetaKthOrder}
\end{equation}

\subsection{Dynamic clustering}
\label{subsection:Dynamic}

Now, in the dynamic clustering, $t_1 = 0$ by definition, and therefore the distribution of $\Delta t = t_m - t_1 = t_m$ is 
\begin{align}
    f_{\rm{D}}(\Delta t \mid m,w ) &= f_{\beta}(t_m \mid m-1,w) \nonumber \\
    &= \dfrac{1}{w} \times \textrm{Beta}\left( \dfrac{\Delta t}{w}; m-1, 1 \right) \nonumber \\
    &= \dfrac{m-1}{w^{m-1}} \times \Delta t^{m-2}.     \label{eq:dynamicDeltaTPDF}
\end{align}
Hence, fixing the multiplicity $m$,  the conditional probability of observing a cluster with $t \leq \Delta t$ is the cumulative distribution of the \autoref{eq:dynamicDeltaTPDF}, i.e.
\begin{align}
    F_{\rm{D}} (\Delta t \mid m) &=\textrm{Prob}(t \leq \Delta t \mid m,w ) \nonumber\\
    &=  \int_{0}^{\Delta t} dt \, f_{\rm{D}} (t | m,w) \nonumber \\
    &= \frac{1}{m} \textrm{Beta}\left( \dfrac{\Delta t}{w}; m, 1 \right) \nonumber \\
    &= \left(\dfrac{\Delta t}{w} \right)^{m-1}.
    \label{eq:dynamicDeltaTCDF}
\end{align}

The discrete multiplicity distribution is provided instead by the truncated Poisson distribution
\begin{equation}
    g_{\rm{D}}(m \mid r,w ) = \frac{\textrm{Pois}(m-1, \mu = rw)}{1-e^{-rw}},    
    \label{eq:truncatedpoisdyn}
\end{equation}
as the cluster contains at least one event by construction of the time window. The joint probability density distribution is then
\begin{align}
    j_{\rm{D}}(m, \Delta t \mid r,w ) &= f_{\rm{D}}(\Delta t \mid m,w ) \times g_{\rm{D}}(m | \, r,w ) \nonumber \\
    &=\frac{re^{-rw}}{1-e^{-rw}} \frac{(r \Delta t)^{m-2}}{(m-2)!}.
\label{eq:jointpdfdyn}
\end{align}

\begin{figure*}[t]
		\includegraphics[width=.95\linewidth]{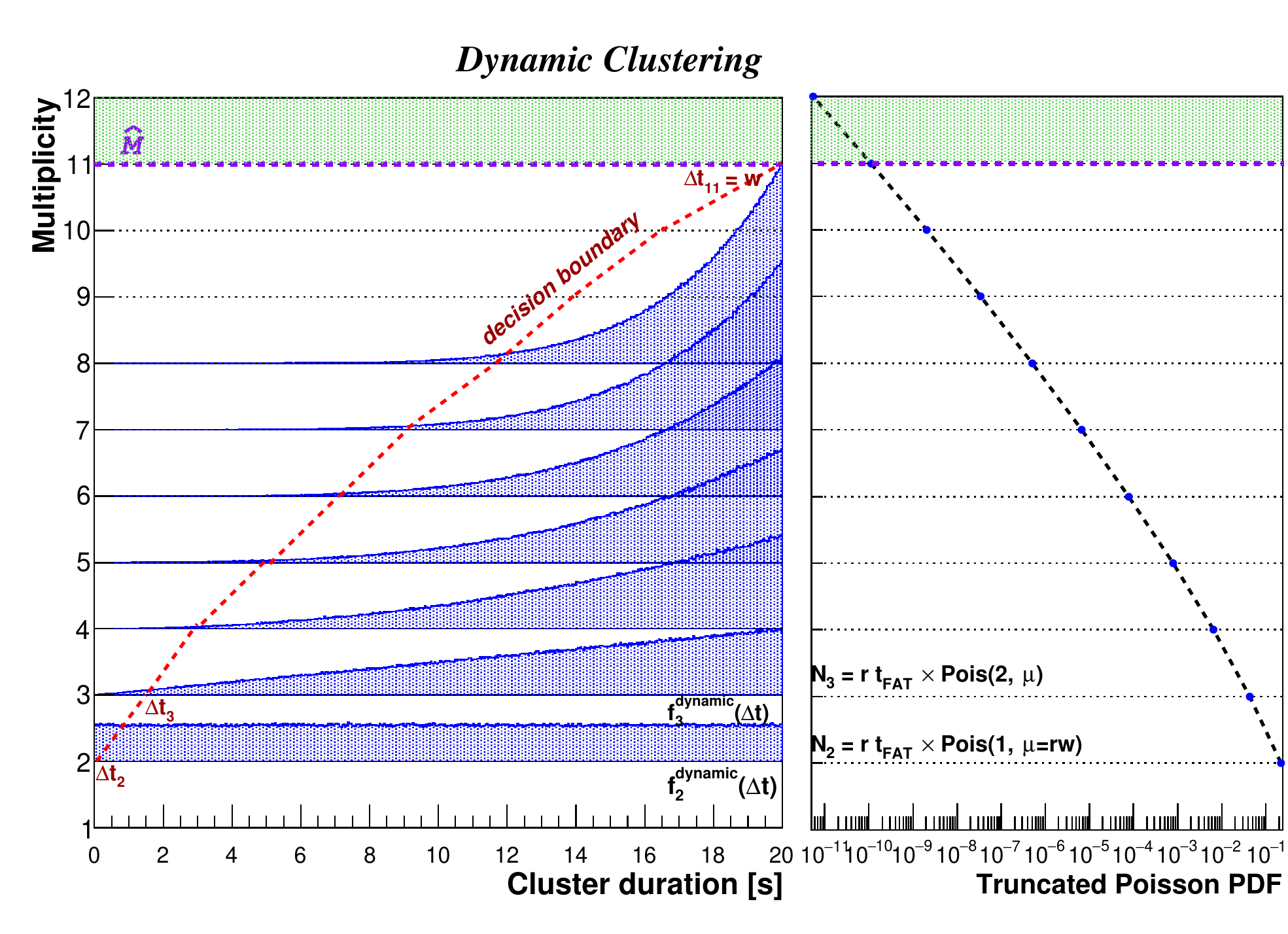}
	\caption{Schematic of the joint probability density distribution as a function of multiplicity $m$ and cluster duration $\Delta t$ for the dynamic clustering. The red dashed line illustrates the decision boundary $\{\Delta t\}_{k=2}^{\infty}$ defined by the new method, ensuring the required false alarm rate. The green region corresponds to clusters selected with the standard method.}
	\label{fig:schematicdynamic}
\end{figure*}

Now, as illustrated in the \autoref{fig:schematicdynamic}, we can construct a decision boundary on the $m - \Delta t$ plane such that the required false alarm rate is satisfied.
It can be defined as a set of values $\{\Delta t_{k}\}$ for $k\geq 2$.

The expected number of clusters with multiplicity $k$ over $t_{FAT}$ is 
\begin{equation}
    N_{k} = rt_{FAT} \times {\rm{Pois}}\, (k-1, \mu = rw), 
    \label{eq:nkcls}
\end{equation}
and the decision boundary must then satisfy the constraint
\begin{equation}
    \sum_{k=2}^{\infty} \alpha_{k} \leq 1  \text{ with } \alpha_{k} :=   N_{k} \times F_{\rm{D}} ( \Delta t_{k} \mid k ).
    \label{eq:constraintsalphadyn} 
\end{equation}

Recalling~\autoref{eq:cumulativeGamma}, if we find $\hat{M}$ such that
\begin{equation}
    \sigma_{\hat{M}} := \sum_{k = \hat{M}}^{\infty} N_{k} = r t_{FAT} \times F_{\gamma}(\hat{M}-1 \mid r,w)< 1,
    \label{eq:Mhatconstraint}
\end{equation}
where a possible choice is the usual multiplicity threshold from the standard imitation frequency method.

Setting $\Delta t_{j} = w$ for $j \geq \hat{M}$, we can rearrange the constraint in \autoref{eq:constraintsalphadyn} as
\begin{align}
    \sum_{k=2}^{\hat{M}-1} \alpha_{k} + \sum_{j=\hat{M}}^{\infty} \alpha_{j}  &\leq 1 \\
    \sum_{k=2}^{\hat{M}-1} \alpha_{k} &\leq 1 - \sigma_{\hat{M}} \\
    \sum_{k=2}^{\hat{M}-1} {\rm{Pois}}\, (k-1, \mu) \times  \left(  \dfrac{\Delta t_{k}}{w} \right)^{k-1}   &\leq 
    \dfrac{1 -  \sigma_{\hat{M}}}{r t_{FAT}} \quad.
\end{align}

Hence, considering the equality, this provides a set of equations that can be exploited to explore the region ${m<\hat{M}}$. In particular, if we introduce the $\{\beta_{k}\}_{k=2}^{\hat{M}-1}$ discrete filter such that
\begin{align}
    {\rm{Pois}}\, (k-1, \mu) \times  \left(  \dfrac{\Delta t_{k}}{w} \right)^{k-1}   &=
    \beta_{k}\times
    \dfrac{1 -  \sigma_{\hat{M}} }{r t_{FAT}} \\
    \sum_{k=2}^{\hat{M}-1} \beta_{k} &= 1.
\end{align}

The final decision boundary is
\begin{equation}
\boxed{
\displaystyle \Delta t_{k} = 
\begin{cases}
w \left [ \beta_{k} \cdot \dfrac{1 - \sum_{j=\hat{M}}^{\infty} N_j}{N_k} \right]^{1/(k-1)} & \text{if $k < \hat{M}$} \\
w  & \text{if $k \geq \hat{M}$}
\end{cases}
}
\end{equation}
where the expected number of clusters with multiplicity $k$ $N_k$ is given in \autoref{eq:nkcls}.

\begin{figure}
    \includegraphics[width=\linewidth]{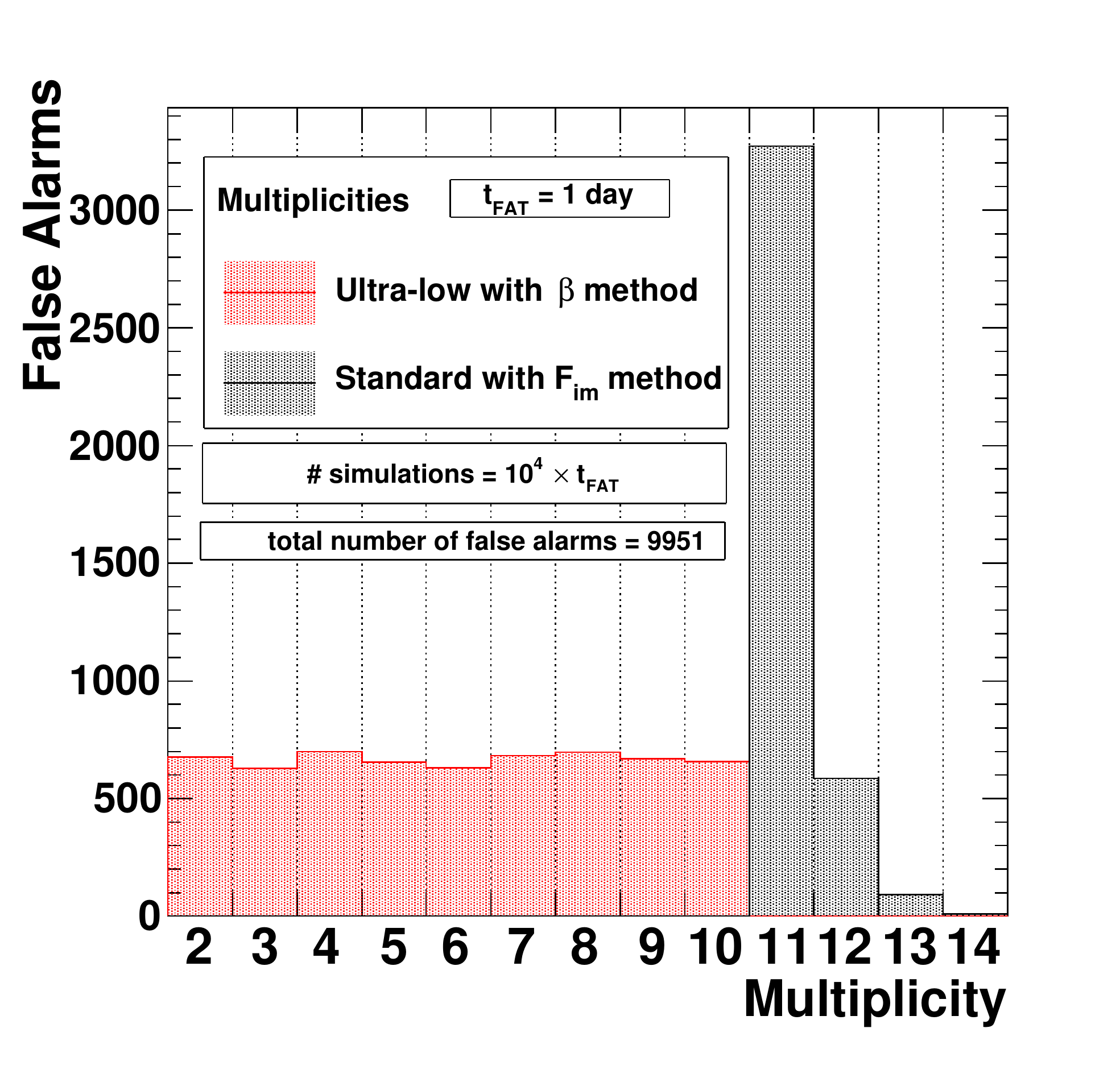}\\[-1em]
    \includegraphics[width=\linewidth]{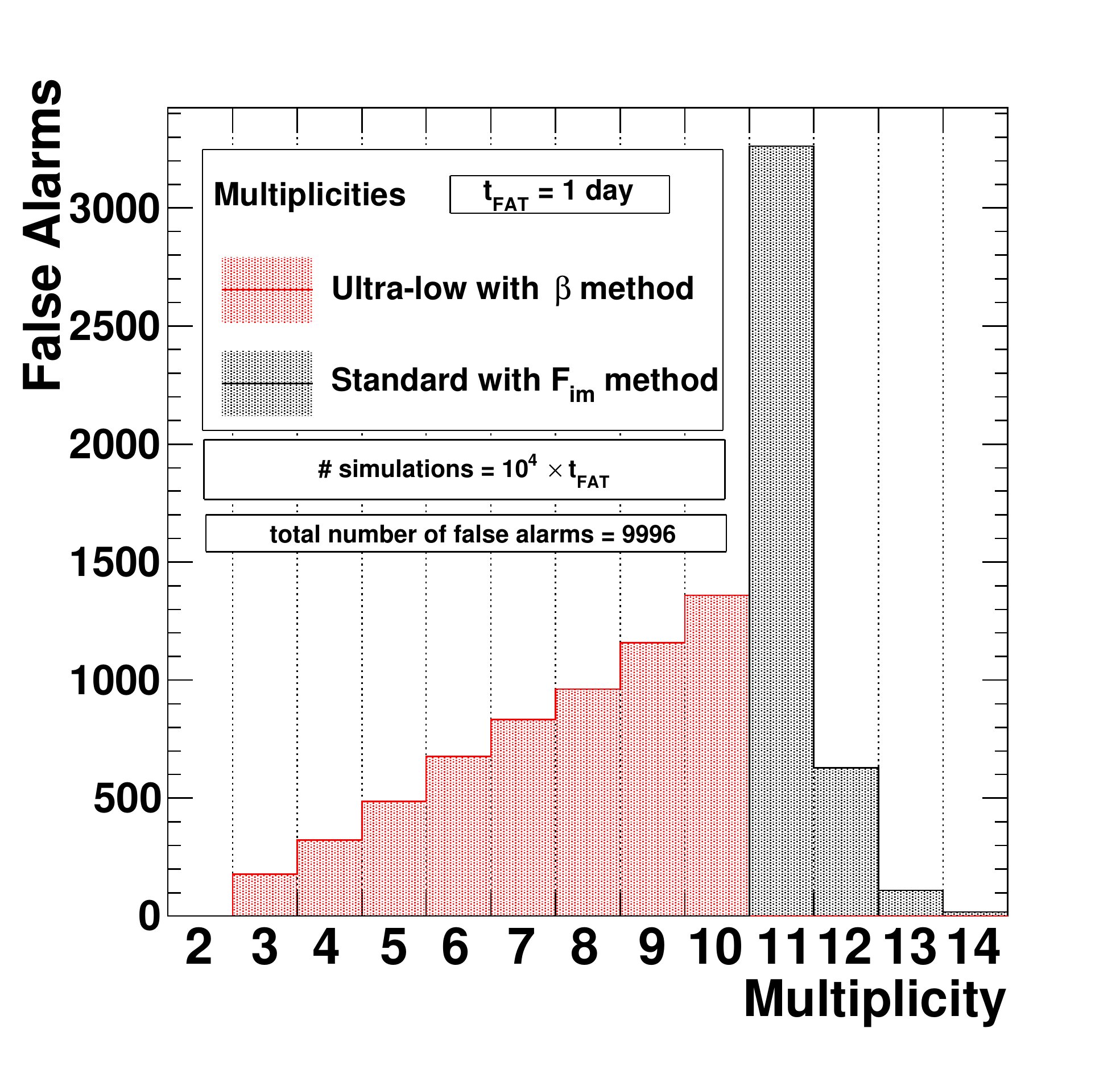}
    
	\caption{(Top) Uniform (Bottom) Ramp. Distributions of false alarms as function of the multiplicity
	applying the uniform (ramp) beta filter, injecting the Poisson-based background at rate $0.1$ Hz with dynamic clustering and performing $10^{4}$ simulations lasting as long as the false alarm time
	under the hypothesis of $t_{FAT} = 1$ day.}
	\label{fig:betafilter}
\end{figure}

Two possible implementations are described in the next paragraphs and illustrated in~\autoref{fig:betafilter}. Nevertheless it is worth emphasizing that the flexibility of this method enables us to set additional application-specific constraints, e.g. discard the multiplicities with associated $\Delta t_{k}$ below the timing resolution of the detector, or penalize the region where non-Poisson background may intervene. 

\subsubsection{Uniform filter} 
If one wants all multiplicities $m<\hat{M}$ to contribute equally, we can use
\begin{align}
    \beta_{k} &= \beta, \quad k \in [2, \hat{M} -1 ] \\
    \sum_{k=2}^{\hat{M}-1} \beta &= 1 \longrightarrow \beta = \frac{1}{\hat{M}-2}.
\end{align}

\subsubsection{Ramp filter} 
If the relative weights scale linearly, the filter is
\begin{align}
    \beta_{k} &= (k-2) \times \beta, \quad  k \in [2, \hat{M} -1 ] \\
    \sum_{k=2}^{\hat{M}-1} \beta_{k} &= 1 \longrightarrow \beta = \frac{2}{(\hat{M}-2)(\hat{M}-3)}.
\label{eq:betafilter}
\end{align}

\subsection{Shifted clustering}
\label{subsection:shifted}

\begin{figure}[ht]
    \includegraphics[width=\linewidth, keepaspectratio]{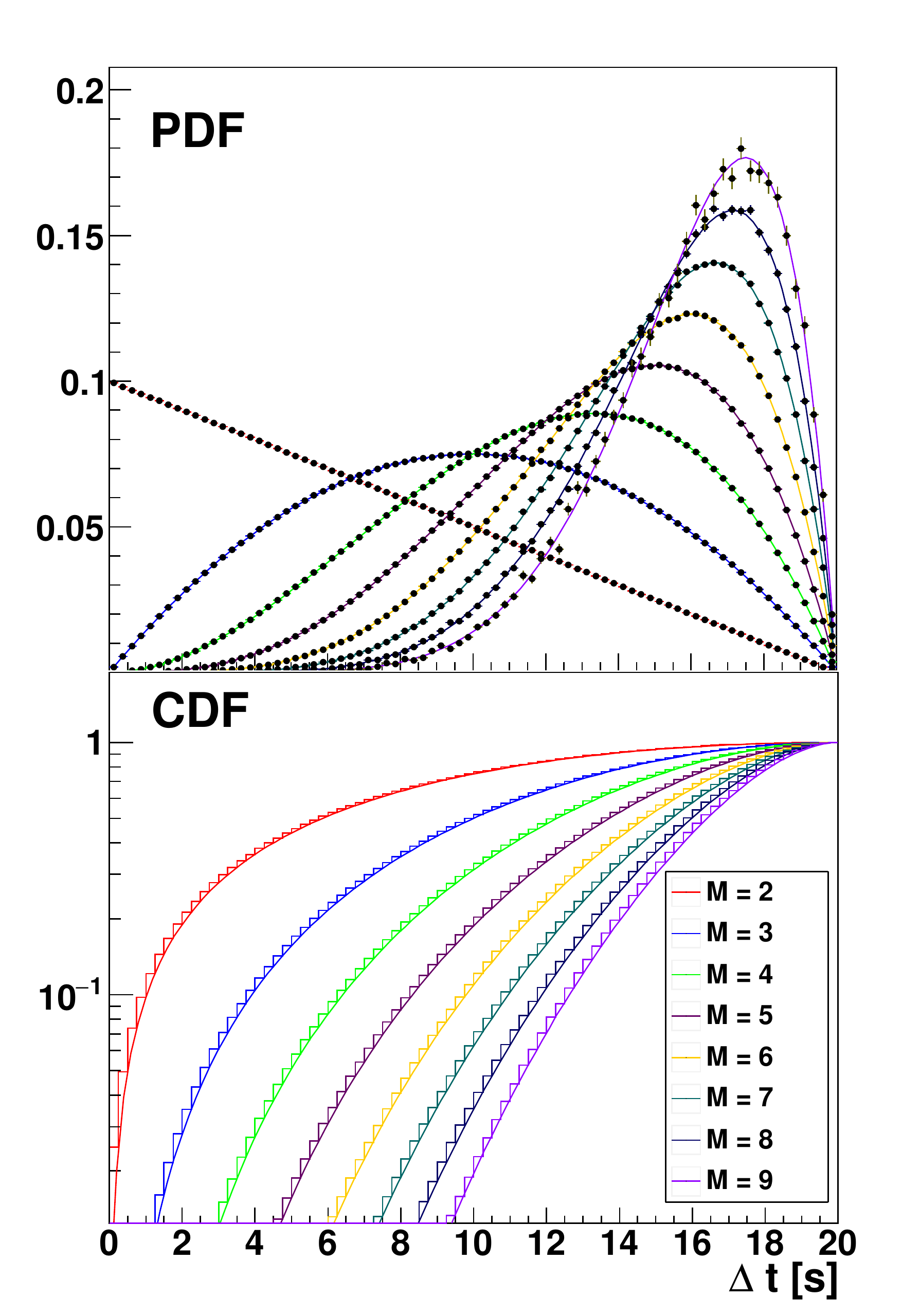}
	\caption{Probability density functions (top) and related cumulative functions (bottom) of the time interval between the first and last events in the shifted clustering, provided by numerical simulations and superimposed (colored lines) with the theoretical curves (equation~\ref{eq:shiftedDeltaTPDF}).}
	\label{fig:betaPDFshifted}
\end{figure}

Conversely, in the shifted clustering,  the probability density function of $\Delta t = t_m - t_1 $ is the sample range of order statistics from the uniform distribution, which, after the specific transformation, is
	\begin{equation}
	f_{\rm{S}}(\Delta t | m,w) = \dfrac{1}{w} \times \textrm{Beta}\left( \dfrac{\Delta t}{w}; m-1, 2 \right).
	\label{eq:shiftedDeltaTPDF}
	\end{equation}
Such distributions have been simulated as reported in~\autoref{fig:betaPDFshifted} and a good agreement is shown. The related cumulative functions (also illustrated in the same figure) are:
\begin{align}
	 F_{\rm{S}} ( \Delta t \, | \, m ) &= \textrm{Prob}(t \leq \Delta t  | m,w ) \\
	 &=  \int_{0}^{\Delta t} dt  \, f_{\rm{S}} ( t | m,w) \nonumber\\
	 &= 	\dfrac{1}{w} m \, (m-1) \int_{0}^{\Delta t} dt  \left ( \dfrac{t}{w} \right)^{m-2}
	 \left ( 1-\dfrac{t}{w} \right) \nonumber\\
	&=  \textrm{Beta}\left( \dfrac{\Delta t}{w}; m, 1 \right) 
	\left [ 1 - \left ( \dfrac{m-1}{m} \right ) \left ( \dfrac{\Delta t}{w}\right ) \right ] .\nonumber 
\end{align}

The discrete multiplicity distribution is again the truncated Poisson distribution, and the joint probability density distribution is
\begin{equation}
j_{\rm{S}}(m, \Delta t \mid r,w ) = \frac{r^{2}e^{-rw} \times (w -\Delta t)}{1-e^{-rw}(1+rw)} \times \frac{(r \Delta t)^{m-2}}{(m-2)!}.
\label{eq:jointpdfshif}
\end{equation}

The expected number of cluster with multiplicity $k$ over $t_{FAT}$ is 
\begin{equation}
N_{k} = N_{W} \times {\rm{Pois}}\, (k, \mu = rw),  
    \label{eq:nkclsshif}
\end{equation}
where $N_{W} = (2 t_{FAT})/w - 1$ is the number of time windows. Now, the same reasoning as for the dynamic case can be applied accordingly:
\begin{equation}
N_{k} \times F_{\rm{S}} ( \Delta t_{k} \, | \, k ) = \beta_{k} \times [1 - \sigma_{\hat{M}}  ] \qquad \text{for} \quad k < \hat{M},
\end{equation}
which can be solved for $\{\Delta t_{k}\}_{k=2}^{\hat{M}-1}$ finding the root between $[0,w]$ of the associated $k$-th degree polynomials.

\section{Performance testing}
\label{sec:performance}

The next step consists in the evaluation of the signal efficiency with the new beta filter method. It is computed as the fraction of simulated SN clusters that pass the new selection as a function of the injected signal multiplicity. However, such efficiency depends on the temporal structure of the signal, which in turn is model-dependent. As suggested in \cite{Casentini2018}, we will use a general-purpose time evolution parametrized as 
\begin{equation}
f_{\rm signal}(t) = e^{-\frac{t}{\tau_{\rm long}}} (1- e^{-\frac{t}{\tau_{\rm short}}}),
\label{eq:SignalShape}
\end{equation}
where $\tau_{\rm short}$ is between \SIlist{10;100}{\milli\second}, and $\tau_{\rm long}$ is $\geq \SI{1}{\second}$. This parametrization enables us to fit not only the SN1987 spectrum (with $\tau_{\rm long} \approx 1$ s) but it can also approximately capture the structure of most of the low-energy neutrino bursts ejected from  similar explosions.

The signal efficiency for the new beta method has therefore been estimated fixing the multiplicity $m_{\rm sig}$, generating $m_{\rm sig}$ signal neutrinos, according to $f_{\rm signal}(t)$ distribution with a random offset $t_0 \in [0,w/2]$ (aiming to avoid the introduction of a systematic error in the shifted case), along with the background events $m_{\rm bkg}$ in the time window $[0, w+w/2]$. Afterwards, the dynamic or the shifted clustering has been applied and one signal injection is considered as ``detected'' if at least one cluster has passed the specific thresholds determined by the decision boundary $\{\Delta t_{k}\}_{k=2}^{\hat{M}-1}$ described in the previous section~\autoref{sec:betafilter}.

The procedure has been iterated $10^{7}$ times, setting $\tau_{\rm short} = \SI{10}{\milli\second}$ and $\tau_{\rm long} = \SI{1}{\second}$, and injecting the Poisson-based noise using LVD-like~\cite{Agafonova2008}, SK-like~\cite{Abe2016}, BOREXINO-like~\cite{Casentini2018}, BAKSAN-like~\cite{Baksan:2019} online background rates. The efficiency of the standard method are also evaluated using the same procedure.

\begin{figure*}[hbtp]
    \includegraphics[width=\linewidth]{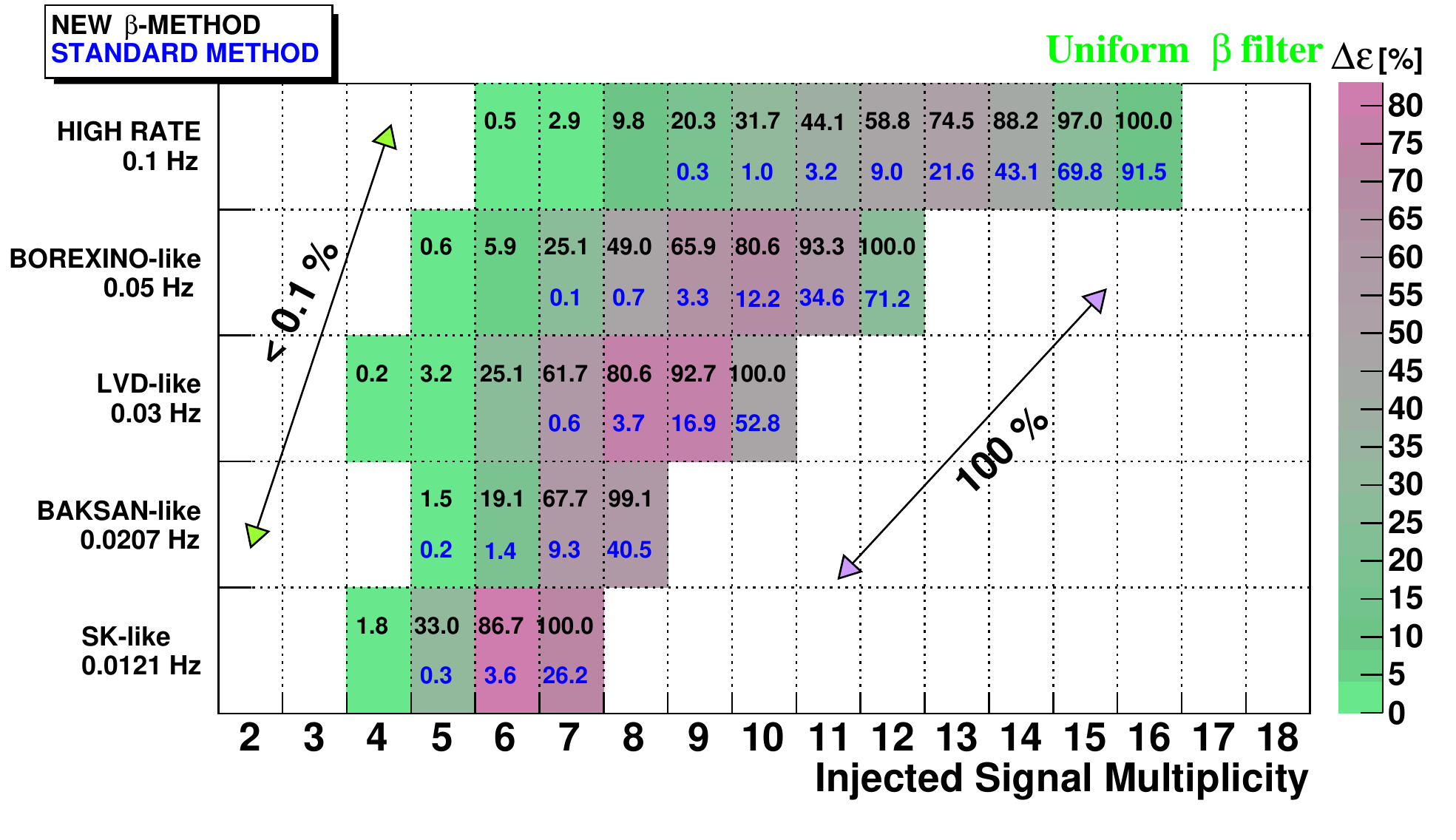}%
    \caption{Signal efficiencies of the standard and the new beta uniform filter methods as a function of injected signal multiplicity, performing the dynamic  clustering, and injecting SN-like signal shape provided by~\autoref{eq:SignalShape} with $\tau_{\rm short} = \SI{10}{\milli\second}$ and $\tau_{\rm long} = \SI{1}{\second}$. The SK-like (\SI{1.2e-2}{\hertz}), BAKSAN-like (\SI{2.0e-2}{\hertz}), low LVD-like (\SI{3.0e-2}{\hertz}), BOREXINO-like (\SI{5.0e-2}{\hertz}) and high-rate (\SI{1.0e-1}{\hertz}) background rates,  have been simulated $10^{7}$ times according to the procedure described in~\autoref{sec:performance}. The bins are colored according to the difference ($\Delta \varepsilon$) between the signal efficiencies of the standard (with blue text) and the beta uniform filter (with black text) methods. Instead, the boxes without the values, have a differential signal efficiency less than $10^{-3}$ or both of them equal to $100 \%$, according to the related colored arrows.  
   \label{fig:signalefficiencyall_dynamic_uniform}}
\end{figure*}

The results are summarized in \autoref{fig:signalefficiencyall_dynamic_uniform} for the dynamic clustering, where the performances of both methods are compared using the beta uniform filter in the new approach (the discrepancies with the ramp filter are small in these cases).

First, the clusters with multiplicity $m_{\rm sig} + m_{\rm bkg} = m \geq \hat{M}$ are selected in the new method as well as in the standard approach. Secondly, the intermediate multiplicities $m \lesssim \hat{M}$ are detected with a much high efficiency, e.g. $\sim 17 \to 93\%$ for $m=9$ in the LVD-like scenario. Lastly, the lowest values of $m$ can be reached with the new beta filter method, whereas they were rejected if only multiplicity is exploited. The two last points could be reformulated in terms of an expansion of the search horizon for SN-like events for all the configurations that have been considered, as it was done in~\cite{Casentini2018} with a different method.

\begin{figure}[!ht]
	\includegraphics[width=\linewidth]{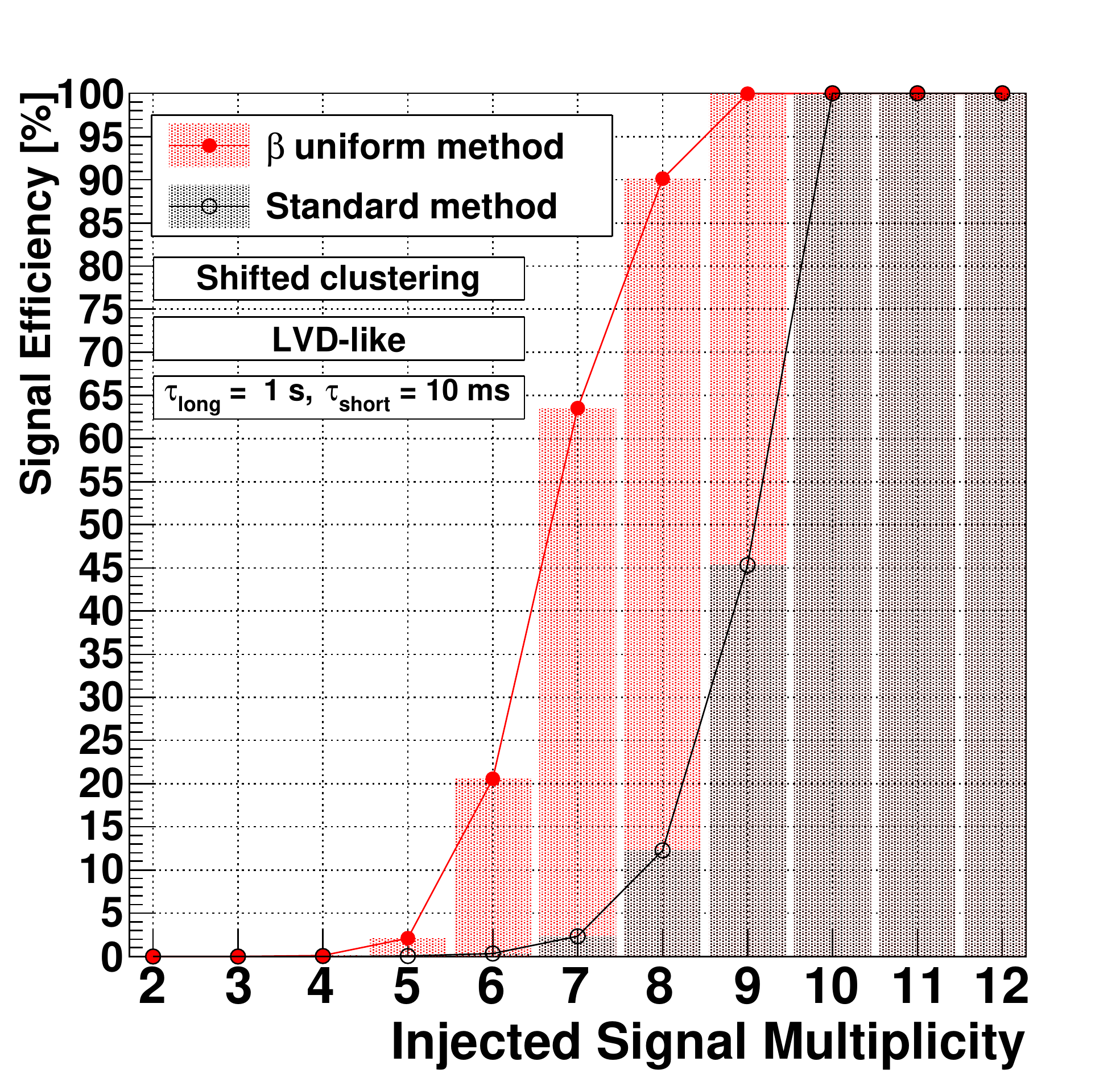} 
	\caption{Signal efficiencies of the standard and the new beta uniform filter methods, for all the relevant low-multiplicities, performing the shifted clustering, and injecting SN-like signal shape provided by~\autoref{eq:SignalShape} with $\tau_{\rm short} = \SI{10}{\milli\second}$ and $\tau_{\rm long} = \SI{1}{\second}$, along with the LVD-like Poisson-based background rate~\cite{Agafonova2008}.}
	\label{fig:effshifted}
 \end{figure}

In \autoref{fig:effshifted}, an example for the shifted clustering with the LVD-like rate, is illustrated, instead.

\section{Conclusions}

Since the standard imitation frequency method is unable to properly exploit the divergent time profiles of the stationary Poisson background with respect to the SN-like signal, we have developed and investigated new beta filter methods which use the interval of arrival time between the first and last events of the cluster as well as the default multiplicity threshold.

The performance was evaluated with the injection of general-purpose SN-like signals along with several rates of Poisson background and the new method has shown important enhancements with respect to standard methods, in particular in the ultra-low multiplicity regime, enabling the extension of the horizon that can be probed in real-time, without a significant upgrade of the monitoring system.

The new beta technique is independent on the nature of the neutrino clusters, therefore it could be suitable for the recognition of other transient astrophysical objects emitting low-energy neutrino bursts, and, more broadly, it could be applied to a wide class of real-time signal-to-noise discrimination processes where sliding-time windows are involved and fixed false discovery rate is required. The only condition for its application concerns the proper identification of correlated backgrounds that are not following a simple Poisson distribution, e.g. spallation events in Super-Kamiokande that are rejected with additional dimension cuts (see \cite{Abe2016}).

In practice, the designed methods could be employed by current low-mass detectors to extend their horizons within the local group, while the planned SN monitor system of the future Hyper-Kamiokande experiment~\cite{protocollaboration2018hyperkamiokande} could also benefit from it: with its \SI{220}{\kilo\tonne} fiducial volume, such new method may indeed allow probing distances up to $\sim\SI{1}{\mega\parsec}$, covering an unexplored region and increasing greatly the rate of detectable CCSNe.

Furthermore, the ongoing SNEWS upgrade (version 2.0) \cite{SNEWS2021} may also be an interesting ground to develop such low-multiplicity techniques, as the strength of the network, originated from its capacity to combine observations from different experiments, reduces the impact from local background sources. 

In conclusion, the gain in sensitivity provided by the new methods would increase the chance to detect a supernova in the coming years and thus probe explosion models (including exotic ones), opening a new window to the exploration of the Universe.

\onecolumngrid
\bibliography{bibliography}

\begin{thebibliography}{}
\expandafter\ifx\csname natexlab\endcsname\relax\def\natexlab#1{#1}\fi
\providecommand{\url}[1]{\href{#1}{#1}}
\providecommand{\dodoi}[1]{doi:~\href{http://doi.org/#1}{\nolinkurl{#1}}}
\providecommand{\doeprint}[1]{\href{http://ascl.net/#1}{\nolinkurl{http://ascl.net/#1}}}
\providecommand{\doarXiv}[1]{\href{https://arxiv.org/abs/#1}{\nolinkurl{https://arxiv.org/abs/#1}}}

\bibitem[{Abe {et~al.}(2016)}]{Abe2016}
Abe, K., {et~al.} 2016, Astropart. Phys., 81, 39,
  \dodoi{10.1016/j.astropartphys.2016.04.003}

\bibitem[{Abe {et~al.}(2018)}]{protocollaboration2018hyperkamiokande}
---. 2018, arXiv e-prints.
\newblock \doarXiv{1805.04163}

\bibitem[{Agafonova {et~al.}(2008)}]{Agafonova2008}
Agafonova, N.~Y., {et~al.} 2008, Astroparticle Physics, 28, 516,
  \dodoi{10.1016/j.astropartphys.2007.09.005}

\bibitem[{Al~Kharusi {et~al.}(2021)}]{SNEWS2021}
Al~Kharusi, S., {et~al.} 2021, New J. Phys., 23, 031201,
  \dodoi{10.1088/1367-2630/abde33}

\bibitem[{Alekseev {et~al.}(1988)Alekseev, Alekseeva, Krivosheina, \&
  Volchenko}]{Alekseev:1988gp}
Alekseev, E., Alekseeva, L., Krivosheina, I., \& Volchenko, V. 1988, Phys.
  Lett. B, 205, 209, \dodoi{10.1016/0370-2693(88)91651-6}

\bibitem[{Antonioli {et~al.}(2004)}]{SNEWS2004}
Antonioli, P., {et~al.} 2004, New J. Phys., 6, 114,
  \dodoi{10.1088/1367-2630/6/1/114}

\bibitem[{Bionta {et~al.}(1987)}]{Bionta:1987qt}
Bionta, R., {et~al.} 1987, Phys. Rev. Lett., 58, 1494,
  \dodoi{10.1103/PhysRevLett.58.1494}

\bibitem[{Casentini {et~al.}(2018)Casentini, Pagliaroli, Vigorito, \&
  Fafone}]{Casentini2018}
Casentini, C., Pagliaroli, G., Vigorito, C., \& Fafone, V. 2018, JCAP, 08, 10,
  \dodoi{10.1088/1475-7516/2018/08/010}

\bibitem[{Fulgione(1996)}]{Fulgione1996}
Fulgione, W. 1996, Nucl. Instrum. Meth. A, 368, 512,
  \dodoi{10.1016/0168-9002(95)00672-9}

\bibitem[{Haines {et~al.}(1988)}]{VanDerVelde:1987hh}
Haines, T., {et~al.} 1988, Nucl. Instrum. Meth. A, 264, 28,
  \dodoi{10.1016/0168-9002(88)91097-2}

\bibitem[{Halim {et~al.}(2020)Halim, Vigorito, Casentini, Pagliaroli, Drago, \&
  Fafone}]{Halim:2019esz}
Halim, O., Vigorito, C., Casentini, C., {et~al.} 2020, J. Phys. Conf. Ser.,
  1468, 012154, \dodoi{10.1088/1742-6596/1468/1/012154}

\bibitem[{Lamoureux(2021)}]{lamoureux2021identification}
Lamoureux, M. 2021, arXiv e-prints.
\newblock \doarXiv{2103.09733}

\bibitem[{Novoseltsev {et~al.}(2020)Novoseltsev, Boliev, Dzaparova, Kochkarov,
  Kurenya, Novoseltseva, Petkov, Striganov, \& Yanin}]{Baksan:2019}
Novoseltsev, Y.~F., Boliev, M.~M., Dzaparova, I.~M., {et~al.} 2020, Astropart.
  Phys., 117, 102404, \dodoi{10.1016/j.astropartphys.2019.102404}

\end{thebibliography}
\bibliographystyle{aasjournal}

\end{document}